\documentclass{article}
\usepackage{codefuse_tech_report}
\usepackage[colorlinks = true,
            linkcolor = blue,
            urlcolor  = blue,
            citecolor = blue,
            anchorcolor = blue]{hyperref}   
\usepackage{microtype}
\usepackage{xurl}
\usepackage{booktabs}
\usepackage{enumitem}
\usepackage{multicol}
\usepackage{CJKutf8}
\usepackage{siunitx}
\usepackage{floatflt}
\usepackage{graphicx}
\usepackage{wrapfig}
\usepackage{authblk}
\usepackage{lipsum}
\usepackage{algorithm}
\usepackage{algorithmicx}
\usepackage{algpseudocode}
\usepackage{subfigure}
\usepackage{multirow}
\usepackage{pifont}  

\algnewcommand{\LeftComment}[1]{\Statex \(\triangleright\) #1}

\usepackage{array}
\usepackage{amsmath}
\usepackage{amssymb}
\usepackage{mathtools}
\usepackage{amsthm}

\usepackage[capitalize,noabbrev]{cleveref}
\usepackage{adjustbox} 

\theoremstyle{plain}

\theoremstyle{definition}

\theoremstyle{remark}

\colmfinalcopy

\usepackage[utf8]{inputenc}
\usepackage[T1]{fontenc}
\usepackage{caption} 
\usepackage{adjustbox} 
\usepackage{fontawesome5}

\usepackage{tcolorbox}
\tcbuselibrary{skins,breakable}

\tcbuselibrary{skins}

\usepackage{color}
\usepackage{xcolor}
\usepackage{soul} 
\usepackage{xspace} 

\usepackage{multirow}
\usepackage{graphicx}

\usepackage{textcomp}
\usepackage{xcolor}
\usepackage{multirow}
\usepackage{subcaption}
\usepackage{enumitem}
\usepackage{tcolorbox}
\usepackage{listings}
\usepackage{booktabs}
\usepackage{graphicx} 
\usepackage{bm}
\usepackage{xspace}
\usepackage{caption}
\usepackage{url}
\usepackage{textcomp}
\usepackage{xcolor}
\usepackage{tcolorbox}
\usepackage{color,xcolor,colortbl}
\usepackage{graphicx}
\usepackage{mathtools}

\usepackage{pifont}      
\usepackage{bbding}
\usepackage{wrapfig}
\usepackage{tcolorbox}
\tcbuselibrary{minted}
\usepackage{subcaption}

\definecolor{mygray}{gray}{.9}

\definecolor{tred}{RGB}{251, 130, 132}
\definecolor{torange}{RGB}{247, 162, 116}
\definecolor{tyellow}{RGB}{251, 218, 140}
\definecolor{tgreen}{RGB}{127, 204, 181}
\definecolor{tblue}{RGB}{89, 177, 215}
\definecolor{insightblue}{RGB}{162, 210, 255}
\definecolor{questionred}{RGB}{255, 175, 204}

\newcommand{\tool}{SWE-Fuse\xspace}
\newcommand{\moduleA}{IFTL\xspace}
\newcommand{\moduleB}{ERLVR\xspace}

\usepackage{etoolbox}

\title{\tool: Empowering Software Agents via Issue-free Trajectory Learning and Entropy-aware RLVR Training}

\vspace{-20pt}
\author{%
Xin-Cheng Wen$^{1}$\thanks{This work was done when Xin-Cheng Wen was a research intern at Ant Group.}
~~Binbin Chen$^{1}$
~~Haoxuan Lan$^{1}$
\\
\vspace{-6pt}
\bf
~~Hang Yu$^{1}$\thanks{Correspondence to: Hang Yu \textless hyu.hugo@antgroup.com\textgreater, Peng Di \textless dipeng.dp@antgroup.com\textgreater and Cuiyun Gao \textless cuiyungao@outlook.com\textgreater.}
~~Peng Di$^{1,\dag}$
~~Cuiyun Gao$^{2, \dag}$

\vspace{2pt}
$^1$Ant Group\ \ \ $^2$The Chinese University of Hong Kong\\
\vspace{2pt}
\hspace{-10pt}\faGithub ~\url{https://github.com/codefuse-ai/xxx}\\
\hspace{-10pt}~~~~~~~~\includegraphics[width=1em,height=1em]{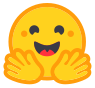} ~\url{https://huggingface.co/datasets/codefuse-ai/xxx}\\
}

\colmfinalcopy 

\begin{document}

\maketitle

\vspace{-2.5em}
\begin{figure}[h!]
    \centering
    \includegraphics[width=0.86\linewidth]{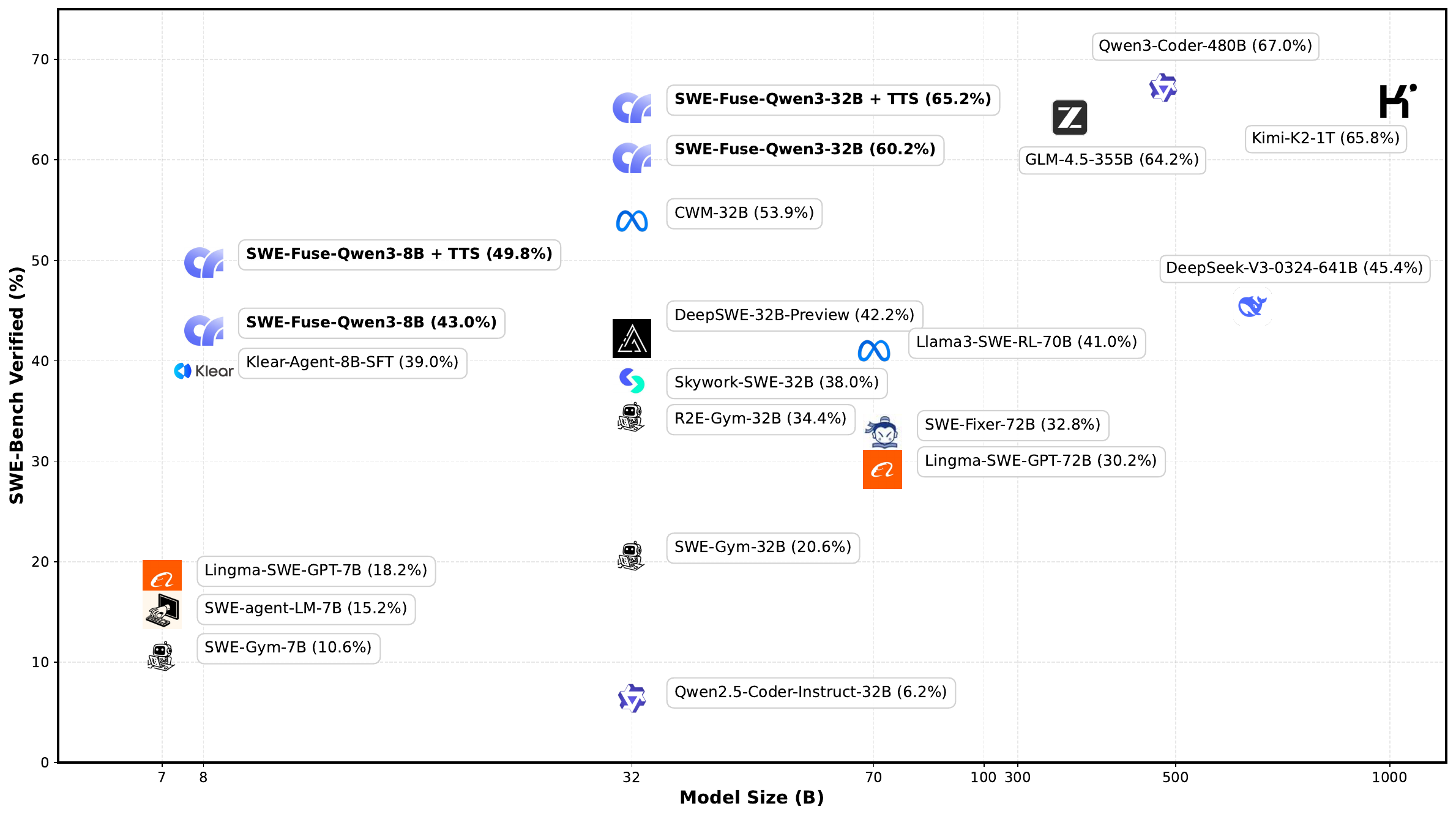}
    \vspace{-1em}
    \caption{Results on SWE-bench Veriefied. 
   \tool ranks first among 8B and 32B models.}
    \label{fig:cover}
\end{figure}
\vspace{-0.7em}

\begin{abstract}
Large language models (LLMs) have transformed the software engineering landscape. Recently, numerous LLM-based agents have been developed to address real-world software issue fixing tasks. 
Despite their state-of-the-art performance, 
Despite achieving state-of-the-art performance, these agents face a significant challenge:
\textbf{Insufficient high-quality issue descriptions.} Real-world datasets often exhibit misalignments between issue descriptions and their corresponding solutions, introducing noise and ambiguity that mislead automated agents and limit their problem-solving effectiveness. 

We propose \textbf{\textit{\tool}}, an issue-description-aware training framework that fuses issue-description-guided and issue-free samples for training SWE agents.
It consists of two key modules:
(1) An issue-free-driven trajectory learning module
for mitigating potentially misleading issue descriptions while enabling the model to learn step-by-step debugging processes; and
(2) An entropy-aware RLVR training module, which adaptively adjusts training dynamics through entropy-driven clipping. It applies relaxed clipping under high entropy to encourage exploration, and stricter clipping under low entropy to ensure training stability.
We evaluate \tool on the widely studied SWE-bench Verified
benchmark shows to demonstrate its effectiveness in solving real-world software problems. Specifically, \tool outperforms the best 8B and 32B baselines by 43.0\% and 60.2\% in solve rate, respectively. Furthermore, integrating \tool with test-time scaling (TTS) enables further performance improvements, achieving solve rates of 49.8\% and 65.2\% under TTS@8 for the 8B and 32B models, respectively.

\end{abstract}

\section{Introduction}\label{sec:intro} 
Large language models (LLMs) have demonstrated substantial capabilities across diverse natural language processing~\cite{ChatGPT, DBLP:journals/corr/abs-2501-12948/DeepSeekR1,DBLP:journals/corr/abs-2402-03300/DeepSeekMath,DBLP:journals/corr/abs-2503-15478/rl1} and software engineering tasks~\cite{DBLP:conf/nips/YangPNY23,DBLP:conf/nips/DingWADTJRNBRX23,DBLP:journals/corr/abs-2308-01861}, including automated bug repair~\cite{DBLP:conf/icse/BouzeniaDP25}, code synthesis~\cite{DBLP:conf/aaai/YeD000JW25}, and vulnerability detection~\cite{DBLP:conf/acl/WenYGXY25/ReVD}. Specialized code-oriented LLMs, such as CodeLlama~\cite{DBLP:journals/corr/abs-2308-12950/codellama}, StarCoder2~\cite{DBLP:journals/corr/abs-2402-19173/starcoder2}, and Qwen2.5-Coder~\cite{DBLP:journals/corr/abs-2409-12186/qwen2.5}, have achieved performance comparable to human developers on numerous function-level programming tasks.

Recently, the evolution of LLM-based software engineering tools has progressed rapidly from basic code autocompletion systems to sophisticated interactive agents capable of autonomous repository-level context processing and end-to-end patch generation~\cite{DBLP:conf/emnlp/ZhangCZKLZMLC23,DBLP:conf/iclr/0003XM24,DBLP:conf/acl/ZhangLLSJ24}. To evaluate these capabilities systematically, SWE-bench~\cite{DBLP:conf/iclr/JimenezYWYPPN24/swebench} was introduced as a benchmark dataset comprising real-world GitHub issues. Initial approaches employed conversational repair systems that leveraged execution feedback for iterative refinement of candidate solutions. Subsequent agentic frameworks, including SWE-agent~\cite{DBLP:conf/nips/YangJWLYNP24/sweagent} and OpenHands~\cite{DBLP:conf/iclr/0001LSXTZPSLSTL25/OpenHands}, enhanced LLMs with integrated development tools, such as terminals, editors, and search functionality for enabling multi-step reasoning across complex codebases. For example, AutoCodeRover~\cite{DBLP:conf/issta/0002RFR24/AutoCodeRover} incorporates dynamic execution traces into the LLM-based repair process, and RepoGraph~\cite{DBLP:conf/iclr/Ouyang0MX0J00025/RepoGraph} employs static code graph representations to support program analysis, though currently limited to Python repositories.
Despite their state-of-the-art performance, these agents still face challenge:

\begin{figure*}[h]
	\centering
	\includegraphics[width=0.9\textwidth]{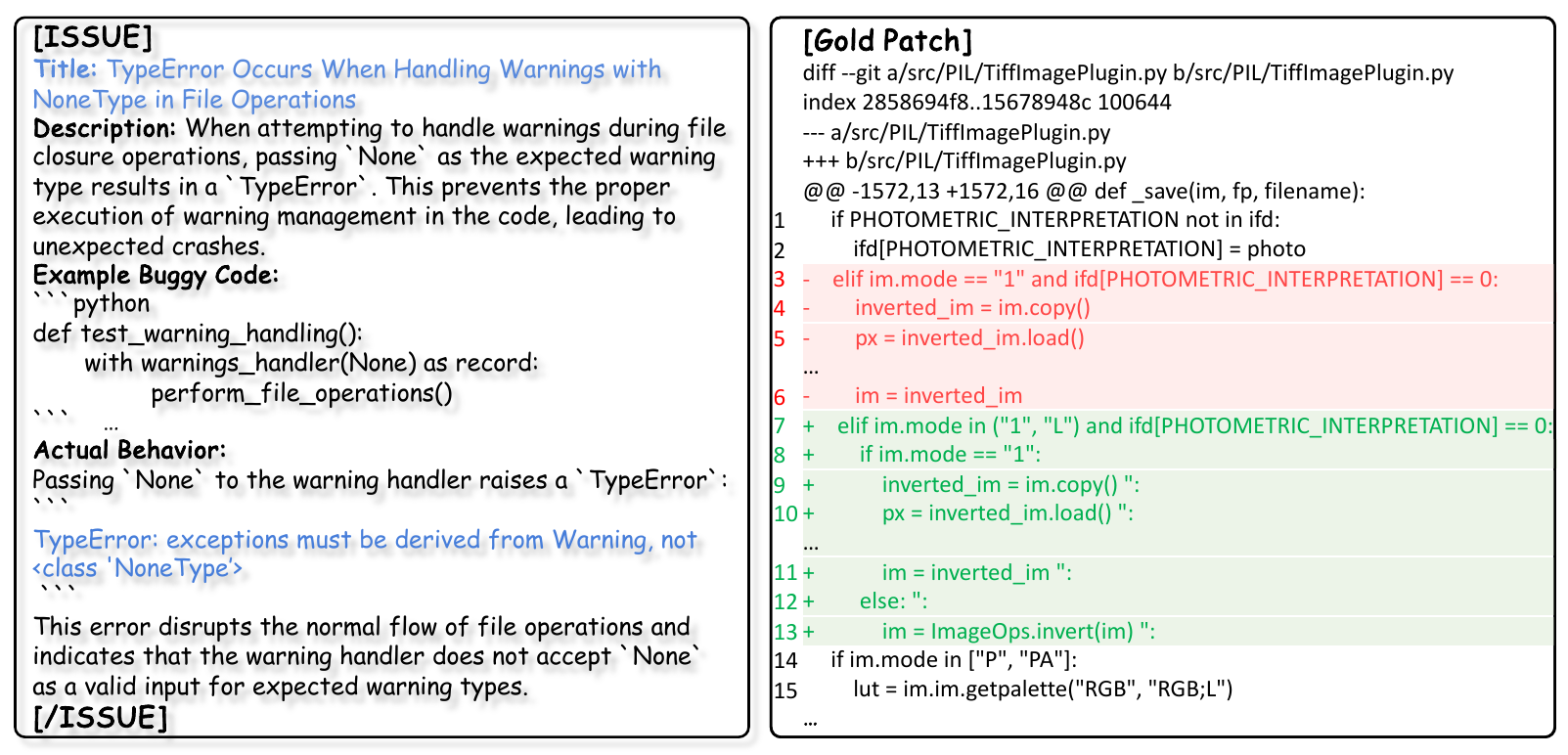}
    \caption{The case in R2E-Gym~\cite{DBLP:journals/corr/abs-2504-07164/r2egym} demonstrating an issue and gold-patch mismatch.}
    \vspace{-0.4cm}
\label{case1}
\end{figure*}

\textbf{Lack of sufficiently high-quality issue descriptions}: 
In SWE-bench, each issue description is authored by domain experts and paired with precisely matched test cases to ensure evaluation accuracy~\cite{DBLP:conf/iclr/JimenezYWYPPN24/swebench}.
However, real-world datasets constructed from authentic scenarios often contain inconsistencies between issue descriptions and their corresponding solutions. These inconsistencies introduce noise and ambiguity that can mislead automated agents, restricting their ability to derive effective solutions~\cite{DBLP:journals/corr/abs-2506-10954/swefac}.
For example, as shown in Fig.~\ref{case1}, 
the issue is about warnings handling in testing, while the patch fixes TIFF image encoding logic. These are two unrelated problems in different parts of the codebase. Specifically,  The issue describes a problem with the warnings handler, when None is passed to \texttt{warnings\_handler(None)}, it causes a ``TypeError: exceptions must be derived from Warning'' (shaded blue). This is a warnings mechanism problem that should be fixed by modifying how \texttt{warnings\_handler} handles None input. However, the gold patch fixes something entirely different—the \texttt{TIFF} image saving logic. It modifies \texttt{PIL/TiffImagePlugin.py} to change how images are inverted (shaded in red). 
Furthermore, high-quality issue-PR pairs are difficult to acquire at scale, and the available datasets remain relatively limited in size.
For instance, in the SWE-smith~\cite{DBLP:journals/corr/abs-2504-21798/swe-smith} dataset, 18,033 samples (30.49\% of the total 59,136 issues) contain empty problem statements, representing a substantial data quality challenge. 

\textbf{Our work.} 
To address these challenges, we propose \textit{\tool}, an issue-description-aware training framework that fuses issue-description-guided and issue-free samples for training SWE agents.
Specifically, our framework enables easier training from base models, with cold-start prior fine-tuning process, achieves stable training with reinforcement learning with verifiable reward (RLVR)~\cite{DBLP:journals/corr/abs-2506-14245/rlvr}, and requires only basic bash commands as tool calls for sandbox environment.

Specifically,  \textit{\tool} consists of two key components:
(1) An issue-free-driven trajectory learning module, which comprises: a multi-step trajectory construction component for generating high-quality multi-turn reasoning-action trajectories, a trajectory data filtering process for ensuring data quality, and issue-free-driven supervised Fine-tuning to mitigate interference from potentially misleading issue descriptions while enabling the model to learn step-by-step debugging processes; and
(2) An entropy-aware RLVR training module, which adaptively adjusts training dynamics through entropy-driven clipping. It applies relaxed clipping under high entropy to encourage exploration, and stricter clipping under low entropy to ensure training stability.

We evaluate \tool on the widely-used SWE-bench Verified~\cite{DBLP:conf/iclr/JimenezYWYPPN24/swebench} benchmark. Experimental results demonstrate that \tool achieves new state-of-the-art performance among open-source models with 32B parameters, resolving 43.0\% of issues with \tool-Qwen3-8B and 60.2\% with \tool-Qwen3-32B.  
Furthermore, integrating \tool with test-time scaling (TTS) enables further performance improvements, achieving solve rates of 49.8\% and 65.2\% under TTS@8 for the 8B and 32B models, respectively.
These results demonstrate that \tool rivals or exceeds the performance of more complex and computationally expensive training methods.

\textbf{Contributions.} The major contributions of this paper are summarized as follows:
\begin{enumerate}

\item 
We propose \textit{\tool}, an issue-description-aware training framework that fuses issue-description-guided and issue-free samples for training SWE agents.
\tool effectively enables reasoning about complex SWE patterns through the issue-free-driven trajectory learning and entropy-aware RLVR training module.

\item 
We introduce the \tool trajectory dataset, comprising 14k validated and correct trajectories. The \tool dataset is constructed from two types of samples: with issue descriptions and issue-free. Issue-free samples help the model avoid influence from noisy descriptions while enabling the LLM to identify problems through systematic debugging.

\item \tool resolves 60.2\% of issues in SWE-bench-Verified, outperforming prior state-of-the-art performance among open-source models with 32B parameters.

\end{enumerate}

\section{Proposed Framework}\label{sec:med}
\begin{figure*}[t]
	\centering
	\includegraphics[width=0.95\textwidth]{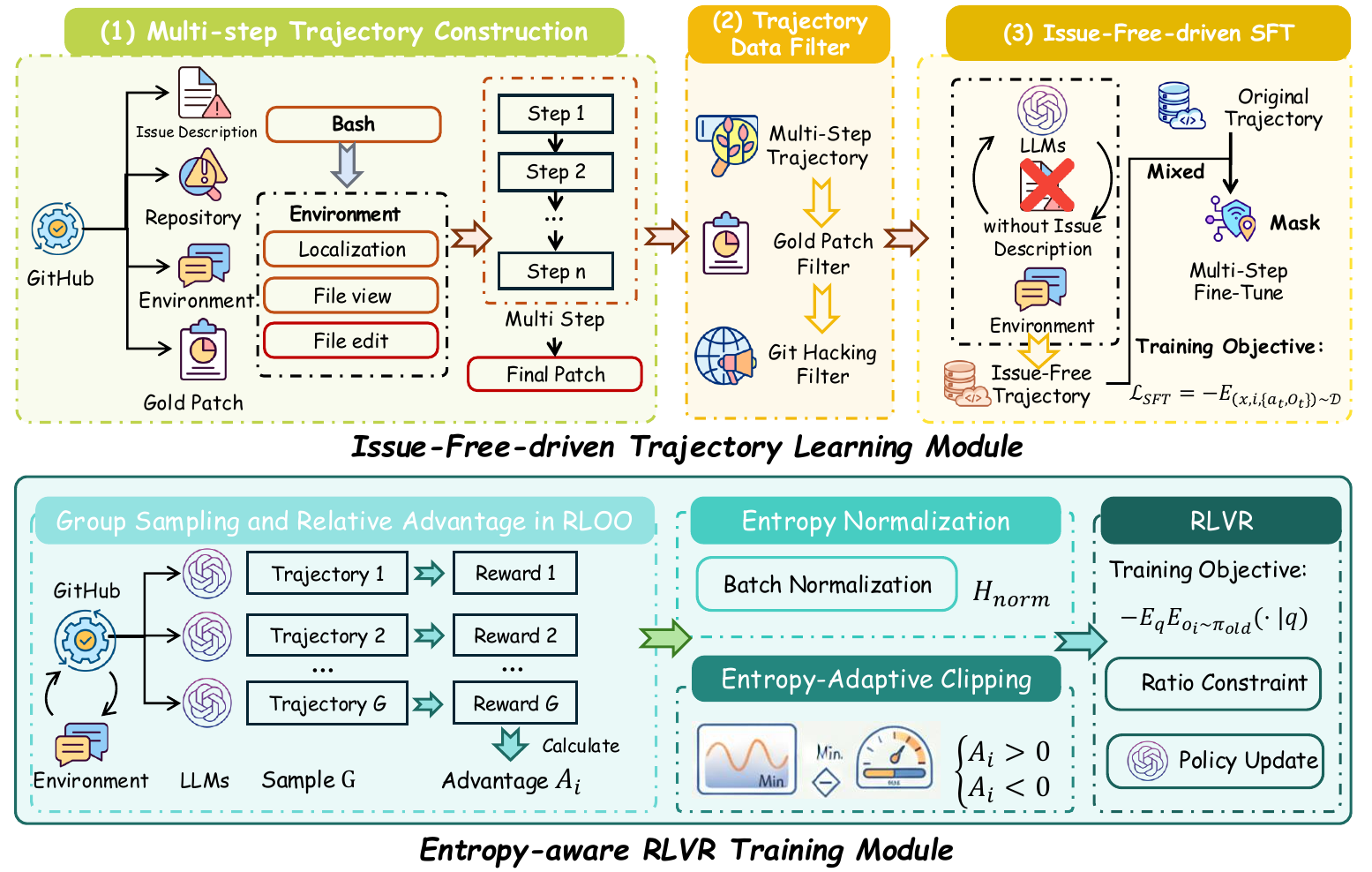}
    \vspace{-0.2cm}
    \caption{The overview of \tool.} 
    \vspace{-0.4cm}
\label{architecture}
\end{figure*}

\subsection{Overview}
Fig.~\ref{architecture} presents an overview of the proposed \tool framework. The primary objective of \tool is to employ a cold-start process that enables the model to first acquire fundamental reasoning capabilities through multi-turn interactions. Concurrently, the model learns to identify and resolve problems by iteratively debugging test case failures across multiple rounds. Guided by RLVR, the model then iteratively selects optimal reasoning steps to accurately complete software engineering tasks, achieving faster convergence under entropy-based guidance.

\textbf{First,} the process begins with the issue-free-driven trajectory learning module, which comprises three components: (1) multi-step trajectory construction, which generates SWE-related reasoning data; (2) trajectory data filtering, which prevents Git exploitation and mitigates the impact of low-quality data; and (3) Issue-Free-driven SFT, which enables the model to complete tasks in issue-free scenarios. \textbf{Subsequently,} \tool employs an entropy-aware RLVR training module. It ensures faster model convergence and more efficient interaction with the execution environment.

\subsection{Issue-Free-driven Trajectory Learning Module}
We propose the issue-free-driven trajectory learning module
for facilitating the initial learning of trajectory reasoning knowledge. 
As shown in Fig.~\ref{architecture},  it mainly contains three phases, including (a) multi-step trajectory construction,  (b) trajectory data filter,  and (c) issue-free-driven SFT training, with details as below:

\subsubsection{Multi-step Trajectory Construction}  

\paragraph{Environment Construction} The environment construction serves two primary purposes: (i) enabling multi-step trajectory generation, and (ii) supporting high-concurrency execution during RLVR phase while enforcing appropriate resource constraints. The construction pipeline comprises three steps:

(1) Repository Collection. \tool builds upon SWE-smith, a dataset of 50,137 instances drawn from 128 executable GitHub repositories. From this corpus, we select more than 33,274 issues with permissive licenses and evidence of active maintenance.

(2) Sandbox Reproduction. We leverage the Docker images provided by SWE-smith~\cite{DBLP:journals/corr/abs-2504-21798/swe-smith}. We retain only repositories that build successfully and pass all sanity checks, resulting in verified base environments for subsequent task construction.

(3) Sandbox Manager Construction. We develop a sandbox manager for scalable and reliable execution, which consists of a management layer for scheduling different environments, and an execution environment layer for single-sandbox execution. (A). a management layer serves as the central scheduling hub responsible for data routing and state management. It communicates upstream via REST APIs to handle user requests, coordinates downstream execution through Tool Call mechanisms to interact with AI sandboxes, and performs file-level data persistence.
(B). An execution environment layer built upon the mini-SWE-agent-plus scaffolding framework, this layer integrates multiple execution engines including bash and Code Interpreter, with extensibility to accommodate additional tool types. 

\paragraph{SWE-Agentic Task Formalization and Trajectory Rollout}

In SWE-bench agentic tasks, we formalize the problem as a sequential decision-making process where an LLM agent $\pi_\theta$ interacts with a software repository environment $\mathcal{E}$ to resolve GitHub issues. Each episode begins with an initial state $s_0$ containing the issue description and repository sandbox, and proceeds through a sequence of interactions until task completion.

\noindent\textbf{ReAct Paradigm.}
At each timestep $t$, the agent executes a two-phase interaction cycle following the ReAct paradigm:
\begin{itemize}
    \item \textbf{Reasoning Phase}: The agent generates an internal reasoning trace $r_t \in \mathcal{V}^*$ to analyze the problem and plan next steps.
    \item \textbf{Action Phase}: The agent selects and executes a tool operation $c_t$ to modify or inspect the repository.
\end{itemize}



\noindent\textbf{Environment Feedback.}
Upon receiving operation $c_t$, the environment transitions to a new state and returns an observation:
\begin{equation}
    s_{t+1}, o_t \sim \mathcal{E}(\cdot \mid s_t, c_t)
\end{equation}
The observation $o_t$ contains execution results (e.g., file contents, test outputs, error messages). The updated history becomes $h_{t+1} = h_t \oplus (r_t, c_t, o_t)$.

\noindent\textbf{Trajectory and Termination.}
A complete trajectory $\tau = (s_0, r_0, c_0, o_0, \ldots, r_T, c_T, o_T)$ ends when: (1) the agent emits a terminal action $c_T = \texttt{submit}$, (2) the step limit $T_{\max}$ is reached, or (3) the token limit is reached. The trajectory receives a binary reward:
\begin{equation}
    R(\tau) = \mathbb{I}[\text{patch}(\tau) \text{ passes all tests in } \mathcal{E}]
\end{equation}
where $\text{patch}(\tau)$ denotes the code changes accumulated throughout the trajectory.

With the ReAct~\cite{DBLP:conf/iclr/YaoZYDSN023/react} paradigm formalized, we now collect expert trajectories for SFT. We adopt the Mini-SWE-Agent-Plus~\cite{DBLP:journals/corr/abs-2511-05951/klear/minisweplus} scaffold with Gemini 3~\cite{Gemini3} as the teacher agent, configuring the maximum number of interaction turns to $T_{\max} = 100$. Gemini 3 offers competitive coding and reasoning capabilities on challenging SWE tasks, making it suitable for generating high-quality demonstration trajectories for distillation. Since Gemini 3 is a closed-source model, we cannot access its internal thought process. Therefore, we explicitly inject a special marker token \texttt{<THOUGHT>} into the trajectory format during collection. Specifically, we modify the interaction template to explicitly separate the reasoning phase:
$h_t \rightarrow \texttt{THOUGHT:} \; r_t \; \textbackslash n \textbackslash n \; \texttt{\textasciigrave\textasciigrave\textasciigrave {bash}} \; c_t \; \texttt{\textasciigrave\textasciigrave\textasciigrave } \;$
This structured format serves three purposes: (1) it makes the reasoning traces explicitly visible in the training data, (2) it teaches the student model to generate explicit thought processes during inference, and (3) it maintains a simple format that facilitates learning for smaller LLMs (less than 32B). While Gemini 3's actual internal reasoning may differ from the verbalized thoughts, this approach provides a reasonable proxy for teaching the student model to reason before acting.

\subsubsection{Trajectory Data Filter}

\paragraph{Preventing Git Exploitation Filtering} It has been recently recognized by the SWE community~\cite{githacking} that LLM agents can unexpectedly exploit git metadata to locate the ground-truth patch by directly inspecting commit logs. 
To address this, we perform the following filtering procedures:
First, in SWE-bench Verified, we remove all commits and log messages dated after the issue creation date to ensure that future fixes remain invisible to the agent. Second, during trajectory collection, we filter out any trajectories containing commands such as ``\texttt{git show}'' or ``\texttt{git log}'' that could expose sanitized repository history to the agent, thereby preventing the model from exploiting git metadata to bypass the intended learning process. In RQ~\ref{subsec:githack}, we provide a detailed empirical analysis demonstrating that our trajectory data and corresponding trained models are not susceptible to such git metadata exploitation, thereby ensuring the integrity and validity of trajectories dataset and \tool.

\paragraph{Rule-based Filtering} Subsequently, rule-based filtering is applied to assess the format and content of the trajectories. The main filtering criteria are as follows: (1) We filter out samples with fewer than 5 interaction rounds. For samples with fewer than 10 rounds, we verify trajectory correctness by evaluating all test cases to ensure validity. (2) Samples lacking intermediate reasoning steps are identified and filtered using regular expressions, as they fail to adequately capture the reasoning process. (3) We enforce a strict response format for bash commands, where tags such as ``$\texttt{\textasciigrave\textasciigrave\textasciigrave {bash}} \; \texttt{\textasciigrave\textasciigrave\textasciigrave}$'' denote the final bash action; samples not conforming to this format are discarded. (4) We ensure that model-generated trajectories contain only English text. Although non-English content may originate from the sandbox environment, we discard all trajectories containing such content to maintain consistency.

\subsubsection{Issue-Free-driven Supervised Fine-tuning}

Due to the limited availability of high-quality issue descriptions, we include a subset of samples without issue descriptions (referred to as "issue-free" samples). For these samples, we provide partial test cases to enable the model to learn through step-by-step debugging. The rationale behind this approach is to mitigate potential noise introduced by imprecise or misleading issue descriptions. As illustrated in Fig.~\ref{case1}, models are often misguided by inaccurate issue descriptions, leading to incorrect exploration trajectories. Specifically, we retain all problem information except the issue description, 
and generate trajectories following the same procedure as for samples with issue descriptions. However, for issue-free samples, we only include successful trajectories where the model correctly resolves the problem.

Then, construct a mixed dataset $\mathcal{D}_\text{mixed} = \mathcal{D}_\text{issue} \cup \mathcal{D}_\text{issue free}$. Each instance in $\mathcal{D}_\text{issue}$ comprises an initial guidance $x$ and issue description $i$. Each instance in $\mathcal{D}_\text{issue-free}$ comprises only the initial guidance $x$ and issue description $i \in \emptyset$, where these samples serve as implicit bug specifications that guide the debugging process without explicit textual descriptions. We utilize the multi-turn interaction trajectories generated in the previous multi-step trajectory construction step. Each trajectory $\tau = (s_0, r_0, c_0, o_0, \ldots, r_T, c_T, o_T)$ consists of alternating agent actions $\alpha_t = \langle r_t, c_t \rangle$ (reasoning $r_t$ and action operations $c_t$) and environment observations $o_t$. 

The model is trained to autoregressively predict the entire decision sequence by maximizing the log-likelihood over complete trajectories. The objective can be expressed as:

\begin{align}
\mathcal{L}_{\text{SFT}} = -\mathbb{E}_{(x, i, \{\alpha_t, o_t\}_{t=1}^T) \sim \mathcal{D}_{\text{mixed}}} \left[ \sum_{t=1}^{T} \log \pi_\theta(\alpha_t \mid x, i, \{\alpha_j, o_j\}_{j<t}) 
\right]
\end{align}

This formulation enables the model to learn both the sequential decision-making policy across multiple interaction rounds.

\subsection{Entropy-aware RLVR Training Module}
RLVR-based methods updates a policy using \emph{relative} credit assignment computed within a group of $G$ sampled completions for the same prompt, thereby avoiding an explicit critic function. A key stabilizer in RLVR implementations is a PPO-style probability-ratio constraint~\cite{DBLP:journals/corr/abs-2510-01555}, which prevents the new policy $\pi_\theta$ from drifting too far from the behavior policy $\pi_{\theta_{\mathrm{old}}}$ in a single update. However, when the policy is uncertain (high entropy), overly tight clipping can impede learning; when the policy is confident (low entropy), overly loose clipping can induce sudden distribution shift and degrade performance~\cite{DBLP:journals/corr/abs-2505-22617}. We therefore propose an \emph{entropy-aware} RLVR training module, including the following four components.

\paragraph{Group Sampling and Relative Advantage in RLOO}
Given a query (prompt) $q$, we sample a group of $G$ SWE-task trajectory outputs $\{o_i\}_{i=1}^{G}$ from the behavior policy $\pi_{\theta_{\mathrm{old}}}(\cdot\mid q)$. Each output $o_i$ is assigned a scalar reward $R_i$ (e.g., task correctness only). Reward leave-one-out (RLOO)~\cite{DBLP:conf/acl/AhmadianCGFKPUH24/rloo} constructs a variance-reduced, sample-specific baseline by taking the mean reward of the \emph{other} group members (i.e., leaving out the current sample). Concretely, for each $i\in\{1,\dots,G\}$, we define the leave-one-out and the corresponding advantage as follows:
\begin{equation}
A_i \;=\; R_i \;-\; \frac{1}{G-1}\sum_{\substack{j=1, j\neq i}}^{G} R_j.
\label{eq:rloo_baseline}
\end{equation}

Compared with the full group-mean baseline used in GRPO~\cite{DBLP:journals/corr/abs-2501-12948/DeepSeekR1}, \tool yields an unbiased estimate of the expected reward baseline~\cite{DBLP:journals/corr/abs-2508-11800} because it removes the self-coupling between $R_i$.
In small-model training paradigm, this reduced coupling often translates into a lower-variance and therefore more stable advantage estimate for  policy optimization where reward variance and sampling noise are typically higher.

\paragraph{Entropy Normalization}
For each sampled output $o_i=(a_{i,1},\dots,a_{i,T_i})$ under query $q$, 
we compute a sequence-level entropy~\cite{DBLP:journals/corr/abs-2506-14758/EntropyPerspective} $H_i$ (e.g., mean token entropy along the trajectory):
\begin{equation}
H_i \;=\; \frac{1}{T_i}\sum_{t=1}^{T_i}\left(-\sum_{a}\pi_\theta(a\mid s_{i,t})\log \pi_\theta(a\mid s_{i,t})\right),
\label{eq:entropy_seq}
\end{equation}
where $s_{i,t}$ denotes the decoding state (prefix) at time step $t$. 
Then, for each sampled sequence-level entropy $H_i$, we  normalize it within the current minibatch:
\begin{equation}
H_{\mathrm{norm},i} \;=\; \frac{H_i-\min_{\mathrm{batch}}(H)}{\max_{\mathrm{batch}}(H)-\min_{\mathrm{batch}}(H)} \in [0,1].
\label{eq:entropy_norm}
\end{equation}
This batch normalization removes scale effects caused by different prompts and action spaces, so that $H_{\mathrm{norm},i}$ measures the \emph{relative} uncertainty of sample $i$ within the current optimization context.

\paragraph{Entropy-Adaptive Clipping}

We propose a dynamic clipping mechanism that adapts the trust-region width at the sample level by jointly considering (i) the policy's uncertainty over the sampled trajectory and (ii) the update direction indicated by the estimated advantage function. 
Furthermore, advantage estimates are susceptible to noise arising from finite batch sizes and stochastic reward signals. Under such conditions, excessively large updates in the negative direction are particularly detrimental, as they may prematurely suppress exploratory behaviors that could prove beneficial, especially when the sign of the advantage estimate is corrupted by noise.

Concretely, we  map $H_{\mathrm{norm},i}$ to a clipping radius $\epsilon_i\in[\epsilon_{\min},\epsilon_{\max}]$ using an adaptive rule:
\begin{equation}
\epsilon_i \;=\;
\begin{cases}
\epsilon_{\min}+(\epsilon_{\max}-\epsilon_{\min})\,H_{\mathrm{norm},i}, & A_i>0,\\[2pt]
\epsilon_{\max}-(\epsilon_{\max}-\epsilon_{\min})\,H_{\mathrm{norm},i}, & A_i\le 0.
\end{cases}
\label{eq:entropy_adaptive_epsilon}
\end{equation}
When $A_i>0$, higher entropy implies greater uncertainty and therefore a larger admissible update region, enabling faster reinforcement of relatively better-than-baseline samples. When $A_i\le 0$, higher entropy triggers a smaller clipping radius, reflecting a conservative stance toward decreasing the probability of a sample under uncertainty and thereby mitigating the risk of over-penalization caused by noisy or spuriously negative advantages.

\section{Experimental setup}\label{sec:setup}
This section describes the experimental setup, including the task formulation and benchmark, the models evaluated, and the metrics used to assess performance.

\subsection{Research Questions}
\begin{wraptable}{r}{0.3\textwidth}
\centering
\vspace{-1cm}
\caption{Statistics of Project Distribution in SWE-bench Verified and Our Subset.}
\label{tab:project_statistics}
\small
\resizebox{\linewidth}{!}{
\begin{tabular}{lrr}
\toprule
\textbf{Project} & \textbf{Total} & \textbf{SubSet} \\
\midrule
astropy/astropy & 22 & 9  \\
django/django & 231 & 92 \\
matplotlib/matplotlib & 34 & 14  \\
mwaskom/seaborn & 2 & 1  \\
pallets/flask & 1 & 1  \\
psf/requests & 8 & 3  \\
pydata/xarray & 22 & 9 \\
pylint-dev/pylint & 10 & 4 \\
pytest-dev/pytest & 19 & 7  \\
scikit-learn/scikit-learn & 32 & 13  \\
sphinx-doc/sphinx & 44 & 17 \\
sympy/sympy & 75 & 30 \\
\midrule
\textbf{Total} & \textbf{500} & \textbf{200}  \\
\bottomrule
\end{tabular}
}
\vspace{-0.3cm}
\end{wraptable}

\begin{enumerate}[label=\bfseries RQ\arabic*:,leftmargin=.5in]
    \item How effective is \tool compared with the state-of-the-art software agents approaches?
    
    \item What is the impact of incorporating different trajectories data during the SFT phase?

    \item How do RLVR algorithm contribute to the performance of \tool?

    
\end{enumerate}

\subsection{Task Formulation and Benchmark}
Our research focuses on real-world software engineering tasks extracted from GitHub issues, encompassing both bug fixes and feature implementations. We employ the widely-adopted SWE-bench Verified benchmark for evaluation.

In RQ1, we utilize the complete SWE-bench Verified benchmark to ensure fair comparison with existing baselines. For the remaining RQs, we carefully curate a representative subset of 200 tasks to balance experimental rigor with computational feasibility. This subset is constructed through stratified random sampling from the full SWE-bench Verified benchmark, preserving the original proportional distribution of tasks across repositories while excluding cases that may cause evaluation inaccuracies due to network-related issues or other external factors. Table \ref{tab:project_statistics} presents a statistical comparison of the task distribution between our subset and the full SWE-bench Verified benchmark.

\subsection{Baselines and Evaluation Metrics}
\begin{wraptable}{r}{0.55\textwidth}
\centering
\vspace{-0.3cm}
\caption{Summary Statistics of the Trajectories Dataset}
\label{tab:dataset_stats}
\resizebox{1.0\linewidth}{!}{
\begin{tabular}{lrrrr}
\toprule
\textbf{Statistic} & \textbf{Total} & \textbf{Mean} & \textbf{Min} & \textbf{Max} \\
\midrule
Valid Trajectories & 14,350 & -& -& -\\
Instances & 14,329 & -& -& -\\
Projects & 111 & -& -&- \\
\midrule
Interaction Rounds & 401,958 & 28.05 & 10 & 98 \\
Token Consumption & 281,938,584 & 19,676.08 & 4,136 & 65,115 \\
\bottomrule
\end{tabular}
}
\end{wraptable}

\paragraph{Baselines} We compare \tool with several widely-used frameworks and models that have been evaluated on SWE-bench Verified. Our selection criterion is based on the availability of source code and generated patches. To ensure fair and accurate comparison, we select both open-source and closed-source models across two parameter scale categories: 7B+ and 30B+. The evaluated models include the Qwen series~\cite{DBLP:journals/corr/abs-2505-09388/qwen3,DBLP:journals/corr/abs-2409-12186/qwen2.5}, GPT series~\cite{openai_o3_o4mini_2025,openai_gpt4o_2024}, Claude series~\cite{anthropic_claude_sonnet_4_5,anthropicClaudeSonnet}, and other state-of-the-art works. These models are deployed across 6 distinct scaffolding frameworks, including MOpenHands~\cite{DBLP:journals/corr/abs-2504-02605/Mopenhands}, OpenHands~\cite{DBLP:conf/iclr/0001LSXTZPSLSTL25/OpenHands}, R2E-Gym~\cite{DBLP:journals/corr/abs-2504-07164/r2egym}, SWE-agent~\cite{DBLP:conf/nips/YangJWLYNP24/sweagent}, Agentless~\cite{DBLP:journals/corr/abs-2407-01489/agentless}, and Mini-SWE-agent-plus~\cite{DBLP:journals/corr/abs-2511-05951/klear/minisweplus}.

\paragraph{Evaluation Metrics}

We evaluate performance on SWE-bench Verified using a single metric: the \textit{issue resolve rate}, defined as the percentage of issues successfully resolved by the generated patches, as verified by test cases. We compare the performance of \tool against baseline results reported in their respective publications and on the official SWE-bench leaderboard. Additionally, we conduct a fine-grained analysis at the repository level to identify all valid patches and quantify the number of issues uniquely resolved by our approach compared to existing baselines.

\section{Experimental Results}\label{sec:result}

\subsection{RQ1: Comparison with SOTA}
To assess the effectiveness of \tool, we compare \tool models against both open-source and closed sourced models on SWE-bench Verified. 

\subsubsection{\tool vs Open-Source Models}
The results summarized in Table~\ref{tab:RQ1} demonstrate that \tool consistently outperforms all open-source models in both the 8B and 32B categories. Specifically, \tool achieves an relative improvement of 9.1\% for 8B models and 11.7\% for 32B models, respectively. Furthermore, \tool represents the only 8B baseline that successfully applies both SFT and RL training. These results indicate that with proper SFT initialization, \tool-8B model effectively learn from multi-turn agentic RL tasks and achieve further performance gains through RLVR, without requiring extensive computational resources for RL training. Furthermore, integration with TTS built upon the \tool allows the performance to be improved even further, achieving a solve rate of 49.8\% and 65.2\% under TTS@8 for the 8B and 32B models, respectively.

\subsubsection{\tool vs Closed-Source Models}
As shown in Table~\ref{tab:RQ1}, \tool demonstrates competitive performance compared to proprietary models. Specifically, \tool outperforms OpenAI-o3 by 1.8\% in resolved rate, while remaining below Claude-4-Sonnet and Claude-4.5-Sonnet. The competitive performance of \tool can be attributed to effective trajectory learning, which enables the model to acquire SWE-specific knowledge more efficiently, allowing a 32B model to achieve performance comparable to models with approximately 1T parameters. However, the remaining performance gap can be attributed to two factors: first, the model size inherently limits knowledge capacity; second, training on a single task leaves substantial room for improvement through multi-task learning.

\begin{table}[t]
\centering
\caption{Performance comparison of \tool and baselines on the SWE-bench Verified. }
\resizebox{0.99\textwidth}{!}{%
  \renewcommand{\arraystretch}{1.4}
  \setlength{\tabcolsep}{14pt}
\begin{tabular}{llcc} 
\toprule
\textbf{Model} & \textbf{Scaffold} & \textbf{Training} & \textbf{Resolve Rate (\%)$\uparrow$} \\
\midrule
\rowcolor{gray!45} \multicolumn{4}{c}{\textbf{Open-Source Models}} \\
\midrule
\rowcolor{gray!15} \multicolumn{4}{l}{\textit{Parameters $\approx$ 7B}} \\
\hspace{1em}Qwen3-8B~\citep{DBLP:journals/corr/abs-2505-09388/qwen3} & OpenHands & - &  7.6 \\
\hspace{1em}SWE-Gym-7B~\citep{swe-gym} & OpenHands & SFT & 10.6 \\
\hspace{1em}SWE-agent-LM-7B~\citep{DBLP:journals/corr/abs-2504-21798/swe-smith} & SWE-agent & SFT & 15.2 \\
\hspace{1em}Lingma-SWE-GPT-7B~\citep{ma2025swe} & SWESynInfer & SFT & 18.2 \\
\hspace{1em}SWE-Mirror-LM-7B~\citep{wang2025swe} & MOpenHands & SFT & 22.8 \\
\hspace{1em}SWE-Dev-7B~\citep{wang2025swed} & OpenHands & RL & 23.4 \\
\hspace{1em}Klear-Agent-8B-SFT~\citep{DBLP:journals/corr/abs-2511-05951/klear/minisweplus} & Mini-SWE-agent-plus & SFT & 39.4 \\

 \midrule
  \rowcolor{blue!8!white}
  \hspace{1em}\tool-8B & Mini-SWE-agent-plus & SFT $+$ RL &  \textbf{43.0} \\
  \rowcolor{blue!8!white}
  \hspace{2em}+\ TTS@8 & Mini-SWE-agent-plus & SFT $+$ RL & \textbf{49.8} \\
\midrule
\rowcolor{gray!15} \multicolumn{4}{l}{\textit{Parameters $\approx$ 30B}} \\ 
\hspace{1em}Qwen3-32B~\citep{DBLP:journals/corr/abs-2505-09388/qwen3} & OpenHands & - & 23.2 \\
\hspace{1em}SWE-Gym-32B~\citep{swe-gym} & OpenHands & SFT & 20.6 \\
\hspace{1em}R2E-Gym-32B~\citep{DBLP:journals/corr/abs-2504-07164/r2egym} & R2E-Gym & SFT & 34.4 \\
\hspace{2em}+\ TTS@16 & R2E-Gym & SFT & 49.4\\
\hspace{1em}SWE-Dev-32B~\citep{wang2025swed} & OpenHands & RL & 36.6 \\
\hspace{1em}Skywork-SWE-32B~\citep{zeng2025skywork} & OpenHands & SFT & 38.0 \\
\hspace{2em}+\ TTS@8 & OpenHands & SFT & 47.0 \\
\hspace{1em}SWE-agent-LM-32B~\citep{DBLP:journals/corr/abs-2504-21798/swe-smith} & SWE-agent & SFT & 40.2 \\
\hspace{1em}DeepSWE-32B-Preview~\citep{deepswe2025} & OpenHands & RL & 42.2 \\
\hspace{2em}+\ TTS@16 & OpenHands & RL & 59.0\\
\hspace{1em}SWE-Mirror-LM-32B~\citep{wang2025swe} & MOpenHands & SFT & 52.2 \\
\hspace{1em}CWM-32B~\citep{copet2025cwm} & Agentless & CPT + SFT + RL & 53.9 \\
 \midrule
 \rowcolor{blue!8!white}
  \hspace{1em}\tool-32B & Mini-SWE-agent-plus & SFT $+$ RL &  \textbf{60.2} \\
 \rowcolor{blue!8!white}
  \hspace{2em}+\ TTS@8 & Mini-SWE-agent-plus & SFT $+$ RL & \textbf{65.2} \\
\midrule
\rowcolor{gray!15} \multicolumn{4}{l}{\textit{Parameters $\ge$ 100B}} \\
\hspace{1em}DeepSeek-V3-0324~\citep{DBLP:journals/corr/abs-2412-19437/DeepSeekV3} & Internal pipeline & - & 45.4 \\
\hspace{1em}GLM-4.5~\citep{DBLP:journals/corr/abs-2508-06471/GLM4.5} & OpenHands & - & 64.2 \\
\hspace{1em}Kimi-K2~\citep{DBLP:journals/corr/abs-2507-20534/kimik2} & Agentless & - & 65.8 \\
\hspace{1em}Qwen3-Coder-480B~\citep{DBLP:journals/corr/abs-2505-09388/qwen3} & OpenHands & - & 67.0 \\
\midrule
\rowcolor{gray!45} \multicolumn{4}{c}{\textbf{Closed-Sourced Models}} \\ 
\midrule
\hspace{1em}OpenAI-o3~\citep{openai_o3_o4mini_2025} & Mini-SWE-agent & - & 58.4 \\
\hspace{1em}Claude-4-Sonnet~\citep{anthropicClaudeSonnet} & SWE-agent & - & 66.6 \\
\hspace{1em}Claude-4.5-Sonnet~\citep{anthropic_claude_sonnet_4_5} & Mini-SWE-agent & - & 70.6 \\
\hspace{1em}GPT-5~\citep{openai_gpt5_2025} & OpenHands & - & 71.8 \\
\hspace{1em}Gemini 3 Pro Preview~\citep{google_gemini3_2025} & Mini-SWE-agent & - & 74.2 \\
\bottomrule
\end{tabular}
}
\vspace{-0.4cm}
\label{tab:RQ1}
\end{table}

\begin{tcolorbox}[width=\linewidth,boxrule=0pt,top=0pt, bottom=0pt, left=1pt,right=1pt, colback=gray!20,colframe=gray!20]
\textbf{Answer to RQ1:} 
\tool achieves the best overall performance across all open-source models in both the 8B and 32B categories, with improvements of 9.1\% and 11.7\% in resolve rate on the SWE-bench Verified dataset. \tool also achieves solve rates of 49.8\% and 65.2\% under TTS@8 for the 8B and 32B models, respectively.

\end{tcolorbox}

\subsection{RQ2: Analysis of Trajectories Data in SFT}
To systematically explore the contribution of trajectory reasoning data within \tool during the SFT phase, we analyze the impact of three key factors on performance: (1) the quantity of generated trajectories to examine the impact of data scaling, (2) varying ratios of samples with issue descriptions versus issue-free samples to investigate whether samples without descriptions benefit the model, and (3) potential Git hacking risks in our generated trajectory data. To balance experimental rigor with computational costs, we use the Qwen3-8B as the base model and evaluate on the representative subset of 200 tasks as shown in Table~\ref{tab:dataset_stats} in all RQ2 experiments.

\subsubsection{Impact of Trajectories Sample Size }

\definecolor{darkgreen}{rgb}{0,0.5,0}
\begin{wraptable}{r}{0.4\textwidth}
\centering
\caption{Impact of trajectories size on \tool performance.}
\resizebox{1.0\linewidth}{!}{
\begin{tabular}{rrr}
\toprule
\textbf{Size} & \textbf{Resolve Rate(\%)$\uparrow$} & \textbf{Resolved Issues$\uparrow$} \\
\midrule
0 & 13.5 & 27/200 \\
\midrule
1k & 24.5\textcolor{darkgreen}{($\uparrow$+11.0\%)} & 49/200 \\
2k & 31.5\textcolor{darkgreen}{($\uparrow$+18.0\%)} & 63/200 \\
4k & 33.5 \textcolor{darkgreen}{($\uparrow$+20.0\%)}& 67/200 \\
8k & 35.0 \textcolor{darkgreen}{($\uparrow$+21.5\%)} & 70/200 \\
\textbf{All} & \textbf{39.0 \textcolor{darkgreen}{($\uparrow$+25.5\%)}} & \textbf{78/200} \\
\bottomrule
\label{RQ2_1}
\end{tabular}
}
\vspace{-0.3cm}
\end{wraptable}

In this experiment, we systematically vary the training data size across multiple orders of magnitude, ranging from 1k to 14k samples, while preserving the respective sampling constraints at each scale. For each data size configuration, we train models using identical hyperparameters and evaluate their performance on the held-out SWE-Bench Verified test set to ensure fair comparison. To maintain consistency across scales, each larger dataset is constructed as a superset of all smaller datasets.

As shown in Table~\ref{RQ2_1}, our experimental results demonstrate a clear positive correlation between training set size and model performance. Specifically, the resolve rate increases monotonically from 13.5\% with zero sample training to 39.0\% when trained on the complete dataset, representing a 2.9× improvement in performance. The model exhibits substantial gains during the initial scaling phase, achieving a 24.5\% resolve rate with only 1k training examples. The marginal improvements gradually diminish as training data increases. This trend suggests that while additional training data consistently improves performance, the model begins to approach diminishing returns beyond 4k to 8k training examples. Nevertheless, even the fully trained model resolves only 78 out of 200 issues (39.0\%), indicating substantial room for improvement and highlighting the challenging nature of automated issue resolution in real-world software engineering contexts. These findings underscore the importance of high-quality training data in few-shot learning scenarios and suggest that data efficiency remains a critical consideration for SWE agents training.

\subsubsection{Impact of Issue Descriptions and Issue-free Samples Ratio}
In this experiment, we investigate the impact of varying the ratio of issue-free samples in the training data. Issue-free samples refer to trajectories that do not contain explicit issue context. We systematically vary the issue-free ratio from 0\% (all samples contain issue descriptions) to 100\% (no issue descriptions included), with intermediate configurations at 25\%, 50\%, and 75\%. For each ratio configuration, we maintain a fixed total training size (4k) and evaluate model performance on the SWE-Bench Verified test set under identical experimental conditions.

\definecolor{darkgreen}{rgb}{0,0.5,0}
\begin{wraptable}{r}{0.4\textwidth}
\centering
\caption{Effect of issue descriptions and issue-free samples ratio on performance.}
\resizebox{1.0\linewidth}{!}{
\begin{tabular}{rrr}
\toprule
\textbf{Ratio(\%)} & \textbf{Resolve Rate(\%)$\uparrow$} & \textbf{Resolved Issues$\uparrow$} \\
\midrule
0  & 33.5 & 67/200 \\
\midrule
\textbf{25} & \textbf{34.0 \textcolor{darkgreen}{($\uparrow$+0.5\%)}} & \textbf{68/200} \\
\textbf{50} & \textbf{34.0 \textcolor{darkgreen}{($\uparrow$+0.5\%)}} & \textbf{68/200} \\
75 & 30.5 \textcolor{red}{($\downarrow$-3.0\%)}& 61/200 \\
100 & 30.5 \textcolor{red}{($\downarrow$-3.0\%)}& 61/200 \\
\bottomrule
\vspace{-0.4cm}
\label{RQ2_2}
\end{tabular}
}
\end{wraptable}

As shown in Table~\ref{RQ2_2}, our results reveal that the resolve rate remains relatively stable when incorporating up to 50\% issue-free samples, maintaining approximately 34.0\% resolve rate (68/200 issues) compared to 33.5\% (67/200 issues) with purely issue-based training. However, performance degrades notably when the issue-free ratio exceeds 75\%, with the resolve rate dropping to 30.5\%, which represents a 10.3\% relative performance reduction compared to the optimal configuration. These findings suggest that while general code modification patterns captured by issue-free samples can provide complementary debug knowledge, issue-specific context remains crucial for effective automated issue resolution. The optimal performance at moderate issue-free ratios (25-50\%) indicates that a balanced mixture of issue and issue-free trajectories may help the model learn both task-specific problem-solving strategies and generalizable test case debug patterns.

\subsubsection{Potential Git Hacking Risks}
\label{subsec:githack}
Docker images of SWE-Bench Verified released prior to September 3, 2025 may contain ground truth commits, potentially enabling ``git hacking'' scenarios. To verify that \tool does not exploit these environmental shortcuts, we conducted a systematic analysis of agent trajectories. We demonstrate that the agent relies primarily on genuine problem-solving capabilities rather than accessing ground truth solutions through git history.

We manually check the sample of agent trajectories across successful resolutions. The analysis reveals minimal instances of git history exploration commands (e.g., git log, git history) in successful task completions. For example, when agents consider git commands, they quickly recognize environmental constraints (e.g., shallow clones, absent .git directories) and pivot to alternative problem-solving strategies, as illustrated by the following agent reasoning:

\begin{tcolorbox}[width=\linewidth,boxrule=0pt,top=-2pt, bottom=-2pt, left=1pt,right=1pt,      colback=blue!5!white, colframe=blue!25!black,]
\textbf{Trajectories Example:} 
... Let's check git log or git blame? I cannot run git commands that require history if it's a shallow clone or if .git is not fully available. Let me verify: ls -la ... 
\end{tcolorbox}
These findings provide strong evidence that our agent's performance stems from learned problem-solving capabilities rather than exploitation of benchmark artifacts.
\begin{tcolorbox}[width=\linewidth,boxrule=0pt,top=0pt, bottom=0pt, left=1pt,right=1pt, colback=gray!20,colframe=gray!20]
\textbf{Answer to RQ2:} 
\tool exhibits clear data scaling benefits, achieves optimal performance with 25-50\% issue-free trajectories that balance task-specific and generalizable debugging patterns, and relies on problem-solving capabilities rather than benchmark exploitation.
\end{tcolorbox}

\subsection{RQ3: Analysis of RLVR}

\begin{figure}[t]
  \centering
  \begin{minipage}{0.3\textwidth}
    \centering
    \includegraphics[width=\textwidth]{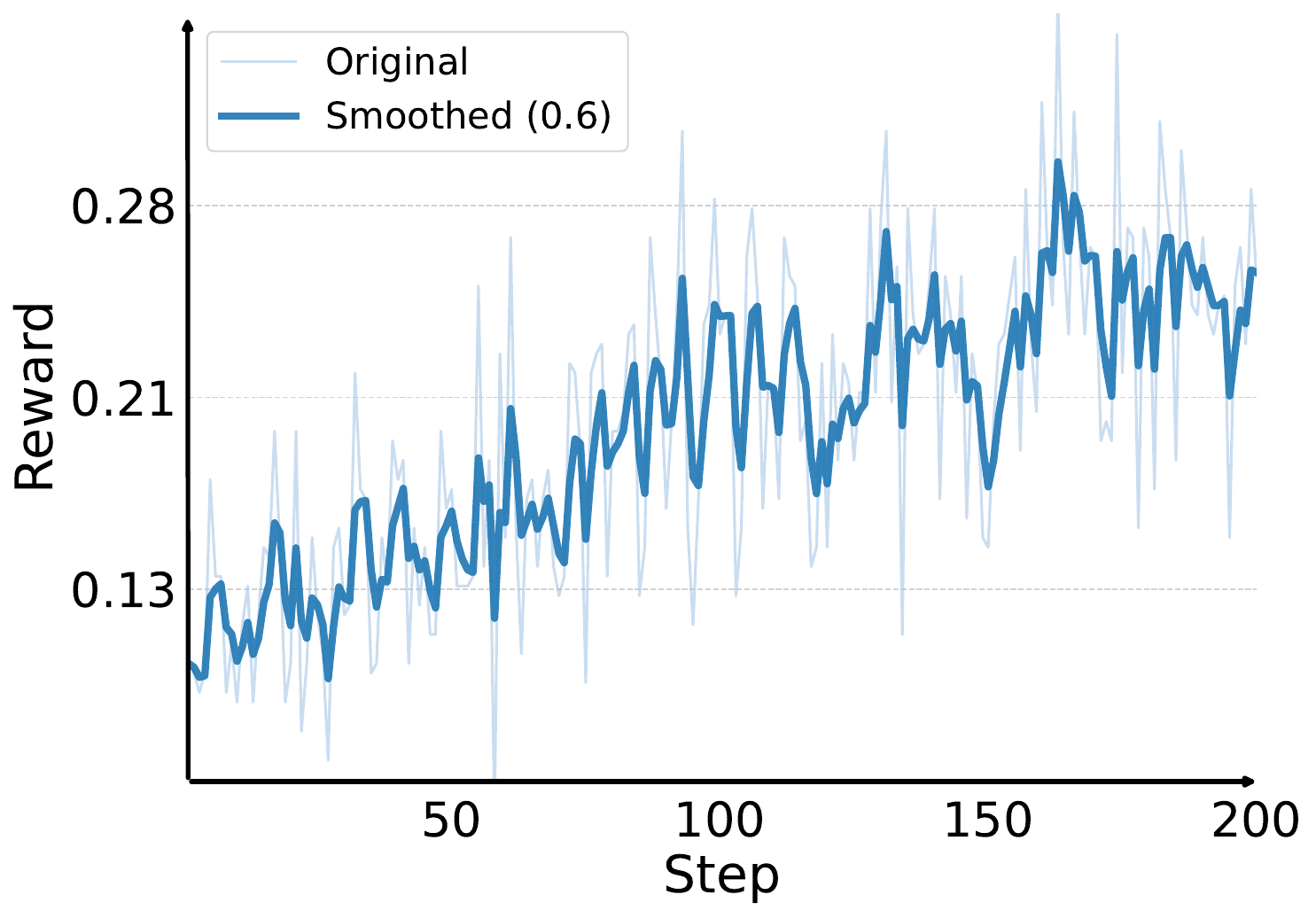}
    \caption{w/o \moduleA Qwen3-8B}
    \label{fig:sub1}
  \end{minipage}
  \hfill
  \begin{minipage}{0.3\textwidth}
    \centering
    \includegraphics[width=\textwidth]{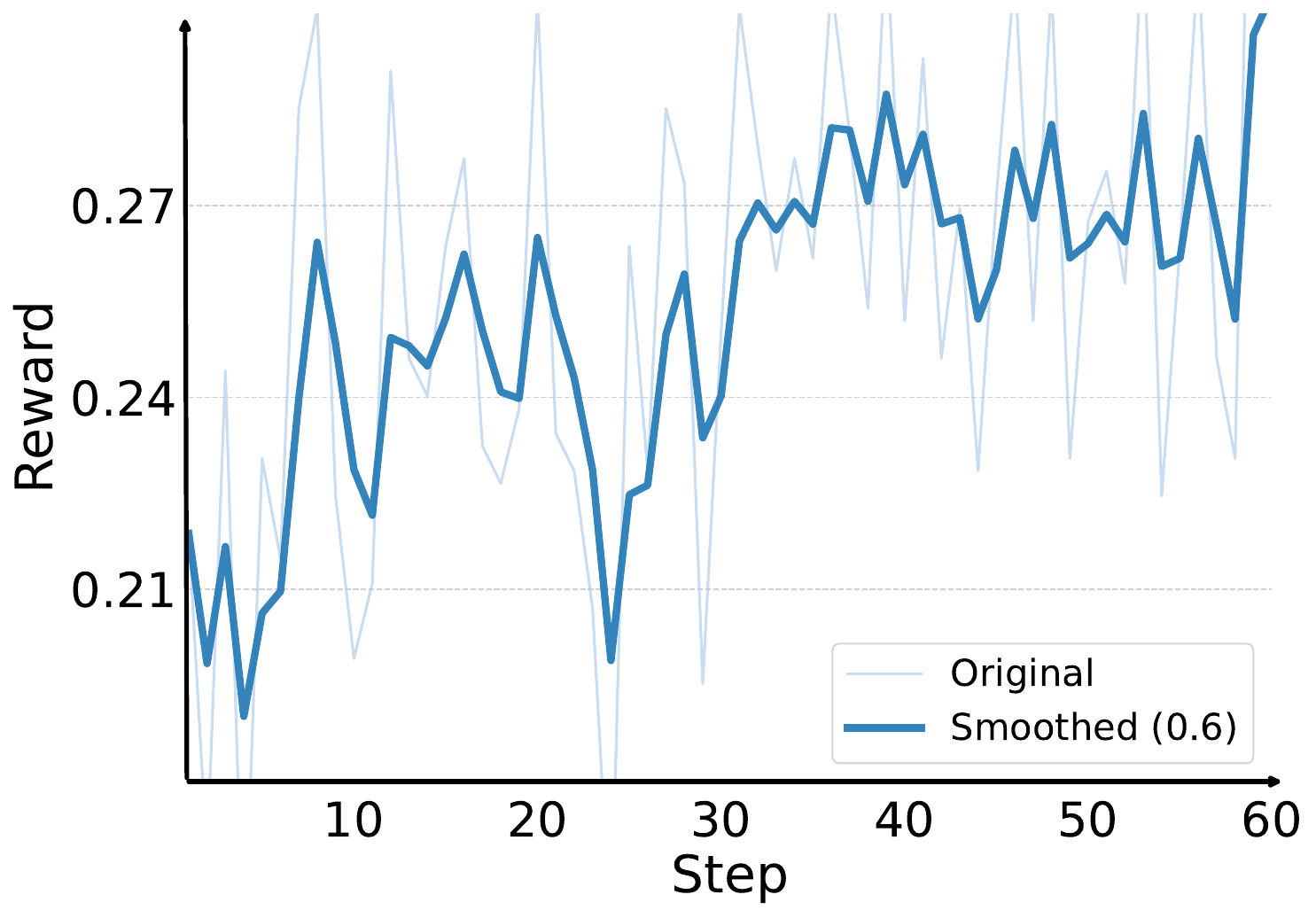}
    \caption{w/ \moduleA Qwen3-8B}
    \label{fig:sub2}
  \end{minipage}
  \hfill
  \begin{minipage}{0.3\textwidth}
    \centering
    \includegraphics[width=\textwidth]{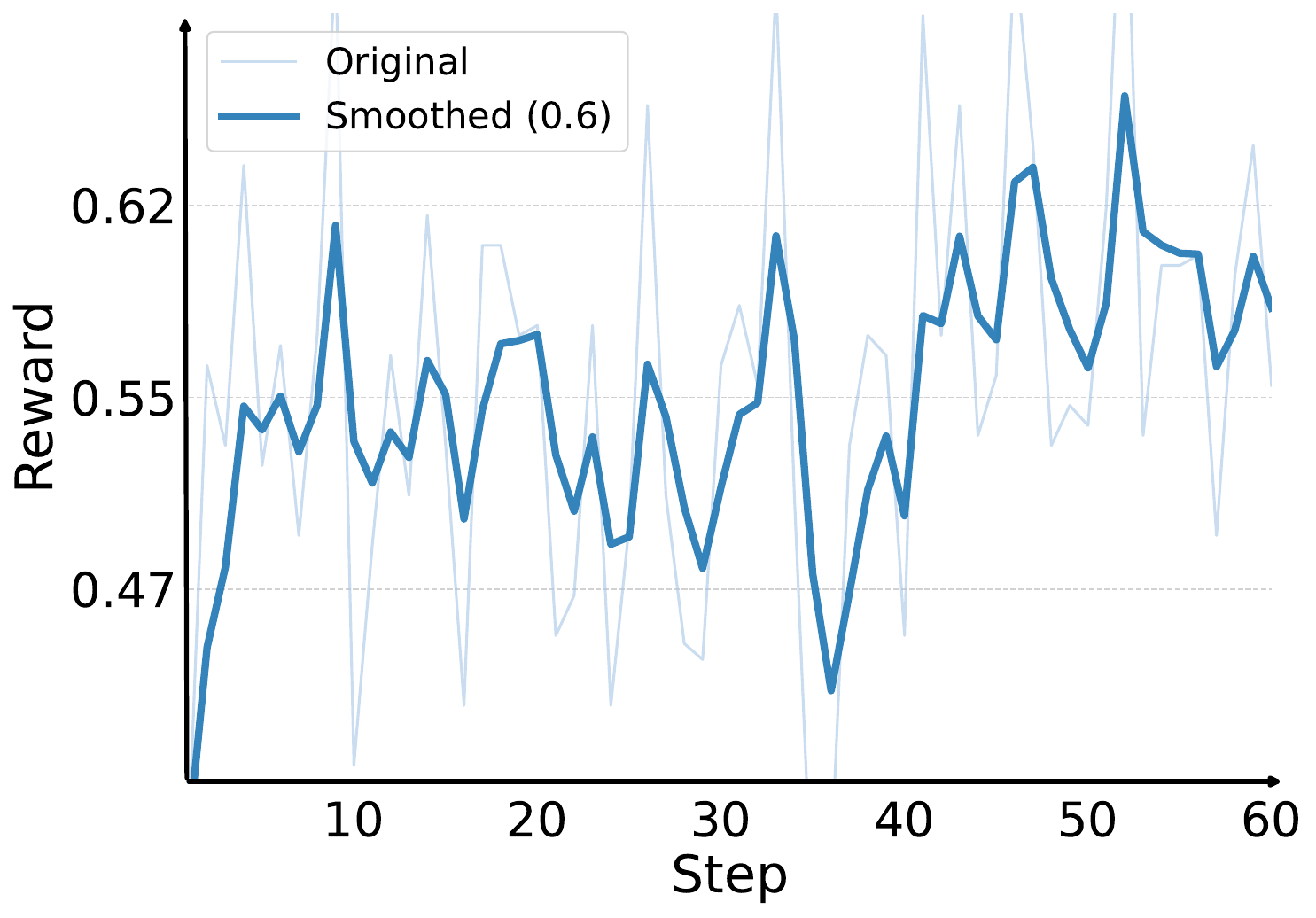}
    \caption{w/ \moduleA Qwen3-32B}
    \label{fig:sub3}
  \end{minipage}
  \caption{The curve of training reward score of \moduleB in \tool.}
  \label{fig:curve}
\end{figure}

To demonstrate the effectiveness of our proposed entropy-aware RLVR training module, we present three training curves in Fig.~\ref{fig:curve}: (1) Qwen3-32B trained without cold-start phase (i.e., w/o \moduleA Qwen3-32B), (2) Qwen3-8B with \moduleA initialization (i.e., w/ \moduleA Qwen3-8B), and (3) Qwen3-32B with \moduleA initialization (i.e., w/ \moduleA Qwen3-8B). 

The reward metric increases progressively throughout training steps across all settings. Notably, Qwen3-32B trained without cold-start phase requires a longer training period compared to SFT-initialized models (more than 200 steps), as it must learn task-specific behaviors from scratch without the benefit of SFT on domain-relevant trajectories. Conversely, SFT-initialized models exhibit rapid performance gains during RLVR and achieve higher ultimate performance. This improvement stems from two complementary mechanisms. First, SFT-based cold-starting mitigates inefficient exploration by biasing the initial policy toward productive action spaces, thereby accelerating training. Second, exposure to curated trajectory data during the SFT phase fundamentally expands the model's capabilities by incorporating task-specific knowledge, which raises the achievable performance ceiling and yields superior final outcomes.

\begin{tcolorbox}[width=\linewidth,boxrule=0pt,top=0pt, bottom=0pt, left=1pt,right=1pt, colback=gray!20,colframe=gray!20]
\textbf{Answer to RQ3:} 
Furthermore, \tool exhibits stable convergence across multiple rounds of agentic RL training on both 8B and 32B models, demonstrating its robustness and scalability across different model sizes.
\end{tcolorbox}

\section{Discussion}\label{sec:discussion}

\begin{figure*}[t]
	\centering
	\includegraphics[width=0.99\textwidth]{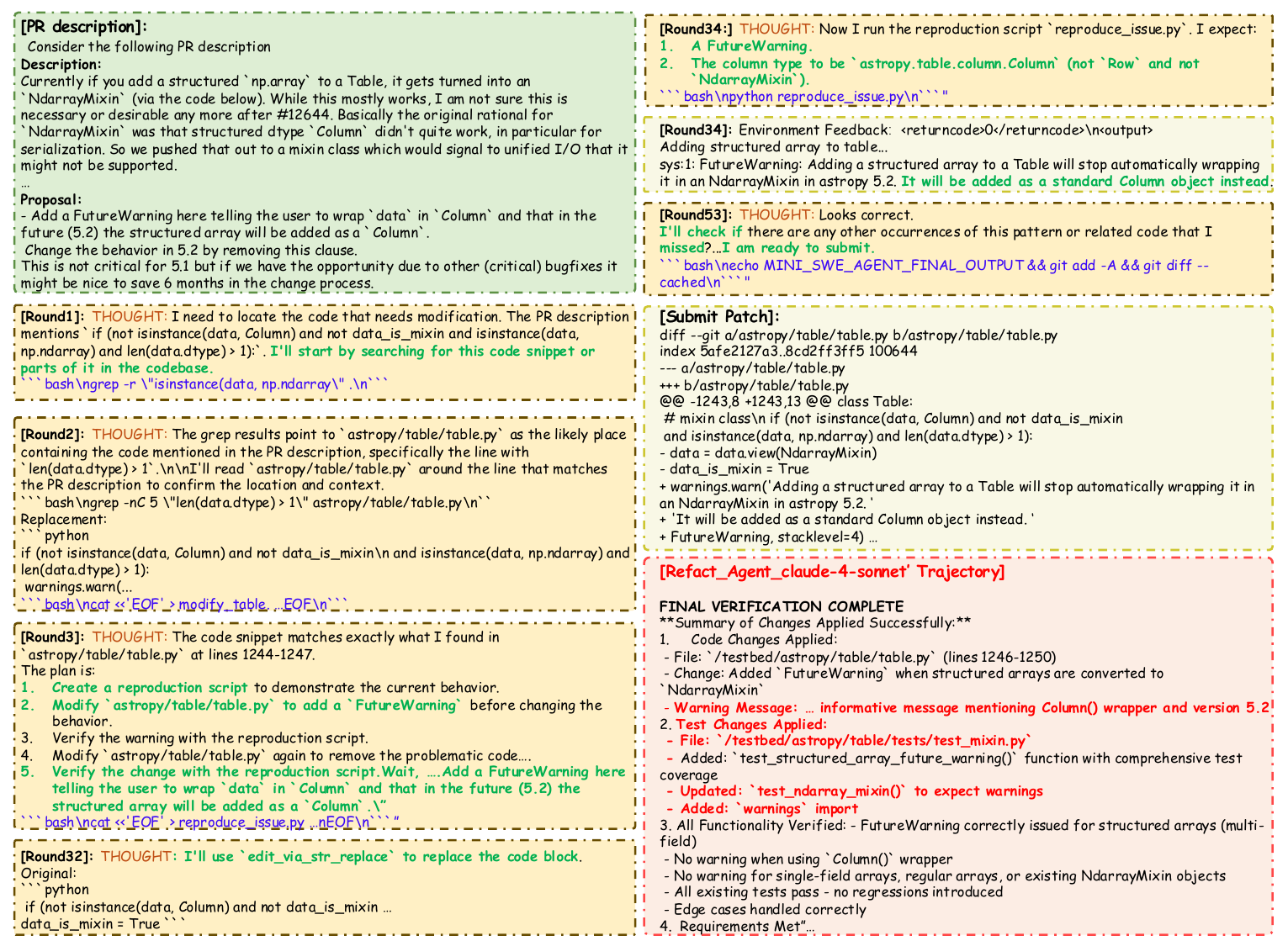}
    \caption{An example case (astropy-13236) from SWE-bench-Verified resolved by \tool.} 
    \vspace{-0.4cm}
\label{fig:case2}
\end{figure*}

\paragraph{Empowering LLMs with step-by-step debugging capabilities by creating reproduction scripts}
After incorporating issue-free trajectories into the training process, \tool effectively learns to create reproduction scripts and generate new test cases to better understand and validate the target pull request descriptions.
As illustrated in Fig.~\ref{fig:case2}, \tool successfully resolved the instance astropy-13236 from SWE-bench-Verified, while Claude-4-Sonnet failed to fix this issue (highlighted in red). Specifically, \tool begins in Round 1 by searching for the relevant code snippet within the codebase. In Round 2, it successfully locates the buggy file \texttt{astropy/table/table.py}. In Round 3, \tool formulates an initial proposal plan, which includes creating a reproduction script, modifying the identified table file, and applying string replacement operations to update the reproduction script. By Round 32, \tool successfully modifies the problematic code block and ultimately submits the correct patch. In this case, \tool utilizes its self-generated reproduction script without altering the original test cases, enabling a progressive debugging process that culminates in successful issue resolution (the submitted patch passes all test cases). In contrast, Claude-4 directly modifies existing test cases when encountering test failures (shown in red-shaded text under "Test Changes Applied"), resulting in an incorrect patch submission. This case study demonstrates \tool's superior effectiveness in real-world issue detection and resolution.

\paragraph{Enhanced reasoning and action through \moduleB for effective multi-turn instruction following}

Beyond the SFT paradigm, we introduce an entropy-aware RLVR training module that enables the model to maintain robust reasoning capabilities within long-context, multi-turn interactions. Specifically, as illustrated in Fig.~\ref{fig:case2}, the pull request description outlines a two-phase proposal: (1) in version 5.1, automatically convert structured arrays to NdarrayMixin, and (2) in version 5.2, remove the \texttt{data.view(NdarrayMixin)} logic to allow structured arrays to become Column objects directly. \tool, at Round 34 during reproduction script execution, correctly addresses both requirements by verifying that: (a) a FutureWarning is issued, and (b) the column type is \texttt{astropy.table.column.Column} (neither \texttt{Row} nor \texttt{NdarrayMixin}). In contrast, Claude-4-Sonnet's final response only implements the first step—adding the FutureWarning when structured arrays are converted to NdarrayMixin—while treating the version 5.2 instruction as merely a warning, ultimately leading to an incomplete issue fix.
The key factor underlying \tool's success is that our modifications strictly align with the pull request's objectives and are validated through reproducible script execution. Throughout the extended multi-turn interaction, \tool comprehensively addresses both the \texttt{add\_column} method and portions of the \texttt{Table(...)} initialization pathway, ensuring precise modifications with verifiable behavioral changes in long-context scenarios.

\section{Related Work}\label{sec:related}

\subsection{Coding LLMs}
LLMs~\cite{kasneci2023chatgpt, chang2024survey, zhao2023surveyllm, naveed2023comprehensive} are pretrained deep learning models predominantly built upon the Transformer~\cite{vaswani2017attention} architecture. In code generation, LLMs are trained on extensive corpora comprising high-quality open-source code repositories and programming documentation~\cite{DBLP:journals/corr/abs-2502-18449/swerl,DBLP:journals/corr/abs-2504-14286/srpo, DBLP:journals/corr/abs-2501-01257/codeforce}. Through this training, LLMs acquire syntactic knowledge across multiple programming languages, learn common programming paradigms, and develop an understanding of the semantic mappings between natural language specifications and code logic~\cite{li2022competition, herrington2003code,herrington2003code,lu2021codexglue}. This enables LLMs to generate executable code from natural language descriptions and perform context-aware code completion and refactoring. Representative code-generative LLMs include Codex~\cite{openai_gpt51_codex_max}, CodeLlama~\cite{DBLP:journals/corr/abs-2308-12950/codellama}, DeepSeek-Coder~\cite{guo2024deepseek}, and Qwen3-Coder~\cite{DBLP:journals/corr/abs-2505-09388/qwen3}. These models have been widely deployed in software engineering tasks such as code completion~\cite{DBLP:journals/tmlr/ShojaeeJTR23/PPOcoder}, test generation~\cite{DBLP:conf/acl/LiuCLZHH0D0025, DBLP:journals/corr/abs-2501-01329}, and automated bug repair~\cite{DBLP:conf/icse/BouzeniaDP25}, demonstrating strong code generation and comprehension capabilities.

\subsection{SWE Agents}
Contemporary agentic agents represent diverse approaches to automating and enhancing software engineering tasks~\cite{DBLP:journals/corr/abs-2511-13646/livesweagent,shrivastava2023repofusion,bairi2023codeplan}.
Several agents primarily focus on streamlining agent-environment interactions. 
For example, Agentless~\cite{DBLP:journals/corr/abs-2407-01489/agentless} introduces a lightweight approach that eliminates the need for complex agent scaffolding while maintaining effective code generation capabilities. SWE-agent~\cite{DBLP:conf/nips/YangJWLYNP24/sweagent} establishes a comprehensive framework that enables agents to interact systematically with software development environments through structured action spaces and observation mechanisms. Building upon this foundation, Mini-SWE-agent~\cite{minisweagent} offers a more compact implementation optimized for resource-constrained scenarios, while Mini-SWE-agent-plus~\cite{DBLP:journals/corr/abs-2511-05951/klear/minisweplus} extends these capabilities with enhanced tool integration. OpenHands~\cite{DBLP:conf/iclr/0001LSXTZPSLSTL25/OpenHands} provides a collaborative multi-agent framework that facilitates concurrent development activities across distributed environments. MOpenHands~\cite{DBLP:journals/corr/abs-2504-02605/Mopenhands} further advances this paradigm by introducing multilingual support and cross-platform compatibility, enabling broader applicability across diverse software ecosystems. Xia et al.~\cite{DBLP:journals/corr/abs-2511-13646/livesweagent} introduce mechanisms for continuous agent improvement, where agents learn from their successes and failures to refine problem-solving strategies on-the-fly, effectively creating a feedback loop that enhances performance over successive interactions.
Collectively, these frameworks serve as foundational scaffolds that structure and orchestrate agent-environment interactions, thereby enabling more systematic and reproducible approaches to automated software engineering.

\section{Conclusion}

We propose an issue-description-aware training framework that fuses both issue and issue-free learning tailored for lightweight models, termed \textit{\tool}
Our framework comprises two key components: (1) an issue-focused trajectory learning module that learns step-by-step debugging processes, and (2) an entropy-aware RLVR training module that adaptively adjusts clipping constraints for stable training. We further release a high-quality trajectory dataset containing 14K trajectories specifically curated for SWE agent training. Experimental results demonstrate that \textit{\tool} achieves state-of-the-art performance for 8B and 32B models on SWE tasks, with solve rates of 43.0\% and 60.2\%, respectively. Integrating TTS@8 further improves performance to 49.8\% and 65.2\%.


\bibliographystyle{colm2024_conference}
\bibliography{custom}

\appendix
\clearpage

\end{document}